\documentclass[12pt,a4paper]{article}

\usepackage{amsfonts,amssymb,amsmath,amsopn,amsthm,graphicx}

\title{Intermixture of extended edge and localized bulk energy levels in
macroscopic Hall systems}
\author{Christian Ferrari and Nicolas Macris}
\date{Institute for Theoretical Physics \\ Ecole Polytechnique F\'ed\'erale \\
CH - 1015 Lausanne, Switzerland}

 \numberwithin{equation}{section}

\newtheorem{thm}{Theorem}

\newtheorem{lem}{Lemma}

\newtheorem{prop}{Proposition}

\newtheorem{hyp}{Hypothesis}

\newcommand{\D}{\,\textrm{d}}
\newcommand{\Le}{\left}
\newcommand{\Ri}{\right}

\newcommand{\N}{\mathbb{N}}
\newcommand{\Z}{\mathbb{Z}}

\newcommand{\R}{\mathbb{R}}

\DeclareMathOperator{\supp}{supp}
\DeclareMathOperator{\dist}{dist}

\DeclareMathOperator{\Tr}{Tr}

\begin{document}
\maketitle

\begin{abstract}
We study the spectrum of a 
random Schr\" odinger operator for
an electron submitted to a magnetic field 
 in a finite but macroscopic two dimensional system of linear dimensions 
equal to $L$. The 
$y$ direction is periodic
 and in the $x$ direction the electron is confined by two 
 smooth increasing boundary potentials. The eigenvalues of the Hamiltonian are
 classified according to their associated quantum mechanical diamagnetic current in the $y$ direction. Here we look at an interval of energies inside the first Landau band of the random operator for
the infinite plane. In this energy interval, with large probability, there exist ${\cal O}(L)$ eigenvalues with
positive or negative currents of ${\cal O}(1)$. Between each of these there
exist ${\cal O}(L^2)$ eigenvalues with
infinitesimal current ${\cal O}(e^{-\gamma B(\log L)^2})$.
We explain what is the relevance of this analysis of boundary diamagnetic
currents to the 
integer quantum Hall effect.
\end{abstract}

\newpage

\section{Introduction}

In this paper we are concerned about boundary currents in the integer quantum Hall
effect, that
occurs in  disordered electronic systems subject to a uniform magnetic
field and confined in a two dimensional 
interface of an heterojunction \cite{PG}.
It was recognized by Halperin that boundary diamagnetic equilibrium currents
play an important role in understanding the transport properties of such systems \cite{H}. 
Later on it was realized
that there is an intimate connection between these boundary currents
and the topological properties of the state in the bulk \cite{FK}, \cite{We}.
Here we will study only diamagnetic currents due to the boundaries, and not
those produced by the
adiabatic switching of an external infinitesimal electric field (as in linear
response theory) which may exist in the bulk.
Many features of the integral quantum Hall effect can be
described in the framework one particle random magnetic Schr\" odinger
operators and therefore it is important to understand their spectral properties 
for finite but macroscopic samples with boundaries. This problem has been approached recently
for geometries where only one boundary is present and the operator is defined
in a semi-infinite region \cite{MMP}, \cite{FGW}, \cite{dBP}.

Here we will take a finite system: our geometry is 
that of a cylinder of length and circumference both equal to  $L$. There are 
two boundaries at $x=\pm\frac{L}{2}$
modelled by two smooth confining potentials $U_\ell(x)$ ($\ell$ for left) and 
$U_r(x)$ ($r$ for right), and we take periodic boundary conditions in the $y$ direction.
These potentials vanish for 
$-\frac{L}{2}\leq x \leq \frac{L}{2}$ and grow fast enough for $|x|\geq \frac{L}{2}$. 
The Hamiltonian is of the form 
\begin{equation}\label{h}
H_\omega=H_0+V_\omega+ U_\ell+U_r
\end{equation}
where $H_0$ is the pure Landau Hamiltonian for a uniform field of 
strength $B$ and $V_\omega$ is a suitable weak random potential produced by 
impurities with $\sup |V_\omega(x,y)|=V_0 \ll B$
(see section 2 for precise assumptions). Before explaining our results it is 
useful to describe what is known about the infinite and semi-infinite
case.
 
In the case of the infinite plane $\R^2$ for the Hamiltonian $H_0+V_\omega$
the spectrum forms ``Landau bands'' contained in  
$\bigcup_{\nu\geq 0} \Le[(\nu+\frac{1}{2})B - V_0,(\nu+\frac{1}{2})B + V_0\Ri]$. It is proved that the band tails have pure point spectrum corresponding to
exponentially localized wavefunctions \cite{DMP1}, \cite{DMP2}, \cite{CH},
\cite{BCH}, \cite{Wa}. 
There are
no rigorous results for energies at the band centers, except for a special model 
where the impurities are point scatterers \cite{DMP3}, \cite{DMP4}. As first shown in \cite{K}
these spectral properties of random Schr\" odinger operators imply that the 
Hall conductivity -- given by the Kubo formula -- considered as a function of the 
filling factor (ratio of electron number and number of flux quanta)
 has quantized plateaux at values equal to $\nu e^2/h$ where $\nu$ is the 
 number of filled Landau levels. The presence of the plateaux is a manifestation of 
Anderson localization while the quantization has a topological origin.
The latter was first discovered in particular situations \cite{TKNN}, and it has been proved for more general
models using non commutative geometry \cite{BES} and the index of Fredholm 
operators \cite{ASS} (see \cite{AG} for a review).

In a semi-infinite  system where the particle is confined in a half plane with
Hamiltonian $H_0+V_\omega+U_\ell$ (here $(x,y)$ belongs to $\R^2$)
the spectrum includes all energies in $\Le[\frac{B}{2}, +\infty\Ri[$. The lower edge of the spectrum is between $\frac{B}{2}-V_0$ and $\frac{B}{2}$ and in its 
vicinity the spectrum is 
pure point (this follows from techniques in \cite{BCH}). For energies 
in intervals inside  the gaps of the bulk Hamiltonian $H_0+V_\omega$ the
situation is completely different. One can show that the average velocity $(\psi,v_y\psi)$
in the $y$ direction of an assumed eigenstate $\psi$ does not
vanish, but since the velocity $v_y$ is the commutator between $y$ and the Hamiltonian, this implies that the eigenstate cannot exist, and that therefore the 
spectrum is purely continuous \cite{MMP}, \cite{F}. In fact Mourre
theory has been suitably applied to prove that the spectrum is purely absolutely 
continuous \cite{FGW},
\cite{dBP}. These works put on a rigorous basis the expectation that, because of the chiral
nature of the boundary currents, the states remain extended in 
the $y$ direction even in the presence of disorder \cite{H}. 
The same sort of analysis shows that if the $y$ direction is  made 
periodic of length $L$, the same 
energy intervals have discrete eigenstates which carry a current that is
${\cal O}(1)$ -- say positive -- with respect to $L$ \cite{FGW}. Furthermore one can show
that the eigenvalue spacing is of order ${\cal O}(L^{-1})$ \cite{M}.

The nature of the spectrum for a semi-infinite system for intervals
inside the Landau bands of the bulk Hamiltonian 
$\bigcup_{\nu\geq 0} \Le[(\nu+\frac{1}{2})B - V_0,(\nu+\frac{1}{2})B + V_0\Ri]$ has not 
yet been elucidated.\\
For the finite system on a cylinder with two boundaries the spectrum consists
of finitely degenerate isolated eigenvalues. In \cite{FM1} the results of 
\cite{MMP}, \cite{FGW} for energy intervals inside the gaps of the bulk Hamiltonian
are extended to the present two boundary system. The eigenvalues can be classified 
in two sets distinguished by the sign of their associated current\footnote{In
principle the physical current is $L^{-1}(\psi,v_y\psi)$, but here we will call
current the average velocity $(\psi,v_y\psi)$.}. These
currents are uniformly positive or uniformly negative with respect to $L$. 
For this result to hold it is important to take the circumference and 
the length of the cylinder both of the order $L$. 

In the present work we study the 
currents of the eigenstates for eigenvalues in the interval 
$\Delta_\varepsilon=\Le]\frac{B}{2}+\varepsilon, \frac{B}{2}+V_0\Ri[$ where $\varepsilon$ is a small positive number 
independent of $L$. We limit ourselves to the first band to keep the discussion simpler. 
The content of our main result (Theorem 1) is the following.
Given $\varepsilon$, for $L$ large enough there is a ensemble of realizations of the random 
potential with probability $1-{\cal O}(L^{-s})$ for which
the eigenvalues of $H_\omega$ can be classified into three sets that we call
$\Sigma_\ell$, $\Sigma_r$ and $\Sigma_b$. The eigenstates of $\Sigma_r$
(resp. $\Sigma_\ell$) have
 uniformly positive (resp. negative) currents with respect to $L$, while those 
 of $\Sigma_b$ have a current of the order of ${\cal O}(e^{-\gamma B(\log L)^2})$. The number of 
eigenvalues in $\Sigma_{\alpha}$ $(\alpha=\ell,r)$ is ${\cal O}(L)$ while that
in $\Sigma_b$ is ${\cal O}(L^2)$. This classification of eigenvalues 
leads to a well defined notion of extended edge and localized bulk states. 
The edge states are those which belong to $\Sigma_\alpha$ $(\alpha=\ell,r)$ and are extended in
the sense that they have a current of order ${\cal O}(1)$. The bulk states are those which belong to 
$\Sigma_b$ and 
are localized in the sense that their current is infinitesimal. 
The energy levels of the extended and localized states are \emph{intermixed} in the same energy interval.
See also \cite{FM3} for a short review on spectral properties of systems defined on a cylinder.

Let us explain the mechanism that is at work. When the random potential is 
removed $V_\omega=0$ in \eqref{h} the eigenstates with
energies away from $\frac{B}{2}$
are extended in the $y$ direction and localized in the $x$ direction at a 
finite distance from the boundaries. Their energies form a sequence of 
``edge levels'' and have a spacing of the order of ${\cal O}(L^{-1})$. When the potential
of one impurity is added to $H_0$  it typically creates a localized bound state with
energy between the Landau levels.   
Suppose now that $i)$ a coupling constant in the impurity potential is \emph{fine 
tuned} as a function of $L$ so that the energy of the impurity level 
stays at distance greater than $L^{-p}$ from the edge levels, $ii)$ the position
of the impurity is at a distance  $D$ from the boundaries.
Then the mixing between the localized bound state and the extended edge
states is controlled in second order perturbation theory by the parameter
$L^{p} e^{-c B  D^2}$. Therefore one expects that bound states of impurities
that have $D\gg (\log L)^{1/2}$ are basically unperturbed 
and have an infinitesimal current. On the other hand bound states coming from impurities
with $D\ll(\log L)^{1/2}$ will mix with edge states.
Note that \emph{even for 
impurities with $D\gg(\log L)^{1/2}$ the coupling constant} (equivalently the impurity level) 
\emph{has to be fine tuned} as a function of $L$.
Indeed, for a coupling constant with a fixed value the energy of the impurity
level is independent of $L$, and surely for $L$ large enough the energy
difference between the impurity and the edge levels becomes much smaller than ${\cal O}(e^{-cBD^2})$.
Remarkably for a random 
potential the absence of resonance is automatically achieved with 
large probability and no fine tuning is needed: this is why localized bulk states
survive. We have analyzed this mechanism 
rigorously for a model (see also \cite{H}) where there are no impurities
in a layer of thickness $(\log L)$ along the boundary. Then the edge levels 
are basically non random and the typical spacing between current carrying 
eigenvalues is easily controlled. Of course it is desirable to allow  for
impurities  close to the boundary but then the edge levels become random 
and some further analysis is needed. However we expect that the same basic
mechanism operates because the typical spacing between edge levels should 
still be ${\cal O}(L^{-1})$.
In connection to the discussion above we mention that
for a semi-infinite system the bound state of an impurity at any fixed distance from the
boundary turns into a resonance. A similar situation has been analysed in \cite{GM}.
 
We note that the spectral region close to $\frac{B}{2}$ that is
left out in our theorem is precisely the one where resonances between edge and
bulk states may occur because edge states become very dense. 
It is not clear what is the connection with the divergence of the localization
length of the infinite system at the band center.

In the present work we have shown that in quantum Hall samples there exist well
defined notions of extended edge states (current of ${\cal O}(1)$) and localized
bulk states (infinitesimal current). Instead of classifying the energy levels
according to their current one could try to use level statistics. We expect that
the localized bulk states have Poissonian statistics whereas the extended edge
states should display a level repulsion. In fact such a strong form of level
repulsion in proved in \cite{M} for energies in the gap of the bulk Hamiltonian
where only extended edge states exist. It is interesting to observe that in the
present situation both kind of states have \emph{intermixed} energy levels. In
usual Schr\" odinger operators (e.g. the Anderson model on a $3D$ cubic lattice)
it is accepted (but not proven) that they are separated by a well defined
\emph{mobilty edge} (results in this direction have recently been obtained
\cite{JL} under a suitable hypotesis). The states at the band edge are localized
in the sense that the spectrum is dense pure point for the infinite lattice and
has Poisson statistics for the finite system \cite{Mi}. At the band center the
states are believed to be extended in the sense that the spectrum is absolutely
continuous for the infinite lattice and has the statistics of the Gaussian
Orthogonal Ensemble for the finite lattice.\\

Others ways of formulating the notion of edge states have been proposed in
differents contexts. In \cite{AANS} the authors consider a clean system with a
novel kind of chiral boundary conditions. The Hilbert space then separate in two
parts corresponding to edge and bulk states. The bulk states have exactly the
Landau energy and the edge states a linear dispertion relation; the distinction
between them being sharp because of the special nature of boundary conditions.
It would be interesting to extend this definition to disordered systems.
Recently in \cite{HS1} (see also \cite{HS2}) another approach has been used in the context of magnetic
billards. The authors study a magnetic billiard with mixed boundary conditions with mixing parameter $\Lambda$ 
interpolating between Dirichlet and Neumann boundary conditions. They look at the sensibility of
the eigenstates and eigenvalues under the variation of $\Lambda$ and define in this
way an edge state as a state that depends strongly on $\Lambda$.
Let us note that our notion of edge state as well as the other ones all share the
feature that an edge state carries a substatial current.\\

The characterization of the spectrum of \eqref{h} proposed here also has a direct relevance
to the Hall conductivity of the many electron (non interacting) system. 
In the formulation advocated by Halperin \cite{H} the Hall conductivity is computed as
the ratio of the net equilibrium current and the difference of chemical
potentials between the two edges. Consider the many fermion state $\Psi(\mu_\ell,\mu_r,E_F)$ obtained by
filling the levels of $H_\omega$ (one particle per state)
in $\Sigma_\ell\cap \Le[\frac{B}{2}+
\varepsilon,\mu_\ell\Ri]$, $\Sigma_r\cap \Le[\frac{B}{2}+
\varepsilon,\mu_r\Ri]$ and $\Sigma_b\cap \Le[\frac{B}{2}+
\varepsilon,E_F\Ri]$ with
$\frac{B}{2}+\varepsilon<\mu_\ell<E_F<\mu_r<\frac{B}{2}+V_0$.
The total current $I(\mu_\ell, \mu_r, E_F)$ of this 
state -- a stationary state of the many particle Hamiltonian -- is given by the sum of the 
individual physical currents of the filled levels (given by $L^{-1}(\psi,v_y \psi)$).
From the estimates \eqref{rt2} and \eqref{rt4} in Theorem 1 
\begin{equation}
\sum_{k} J_k^\ell + \sum_{k} J_k^r +
\sum_{\beta} J_\beta =
\sum_{k}
J_{0k}^\ell + \sum_{k} J_{0k}^r + {\cal O}(e^{-(\log L)^2}L^2)
\end{equation}
and from \eqref{HF} we get
\begin{equation}
\frac{1}{L}\sum_{k} J_{0k}^r = \frac{1}{2\pi} \int_{\frac{B}{2}+\varepsilon}^{\mu_r} \D E +
{\cal O}(L^{-1})
\end{equation}
\begin{equation}
\frac{1}{L}\sum_{k} J_{0k}^\ell = \frac{1}{2\pi} \int_{\mu_\ell}^{\frac{B}{2}+\varepsilon} \D E +
{\cal O}(L^{-1})
\end{equation}
It follows that to leading order 
\begin{equation}\label{SH}
I(\mu_\ell,\mu_r,E_F)\simeq \frac{1}{2\pi} (\mu_r-\mu_\ell) \; .
\end{equation}
In \eqref{SH} the Hall conductance is equal to one (this is because we have considered
only the first band). When  $\mu_\ell$ and $\mu_r$ vary the density of particles in the state 
$\Psi(\mu_\ell,\mu_r,E_F)$ does not change
since the number of levels in $\Sigma_\alpha$ $(\alpha=\ell,r)$ is of order ${\cal O}(L)$. However if $E_F$ is increased
the particle density (and thus the filling factor) increases since the number of levels in 
$\Sigma_b$ is of order ${\cal O}(L^2)$, but the Hall conductance does not change and hence has a plateau. 
In other words the edge states contribute to the Hall conductance but not to
the density of states of the sample in the thermodynamic limit.

In a more complete theory one should also take in account currents possibly
flowing in the bulk due to the adiabatic switching of an external electric
field, an issue that is beyond the scope of the present analysis.
A related problem is the relationship between the conductance in the present
picture, defined through \eqref{SH}, and the one using Kubo formula (see 
\cite{KRS1}, \cite{KRS2}, \cite{EG}).

The precise definition of the model and the statement of the main result (Theorem 1) are the subject of the next section. 

\section{The Structure of the Spectrum}\label{ch2}

 We consider the family of random Hamiltonians \eqref{h}
acting on the Hilbert space 
$L^2(\R\times [-\frac{L}{2},\frac{L}{2}])$ 
with  periodic boundary conditions along $y$, $\psi(x,-\frac{L}{2})
=\psi(x,\frac{L}{2})$.
 In the Landau gauge the kinetic term of \eqref{h} is
\begin{equation}\label{1}
H_0=\frac{1}{2}p_x^2 + \frac{1}{2}(p_y-Bx)^2
\end{equation}
and has infinitely degenerate Landau levels
$\sigma(H_0)=\Le\{(\nu+\frac{1}{2})B; \nu\in \N\Ri\}$. We will make extensive
use of explicit point-wise bounds, proved in Appendix A, on the integral kernel  of
the resolvent $R_0(z)=(z-H_0)^{-1}$ with periodic boundary conditions along $y$.

The confining potentials modelling the two edges at $x=-\frac{L}{2}$ and
$x=\frac{L}{2}$ are assumed to be strictly monotonic, differentiable and 
 such that
\begin{eqnarray}
c_1|x+\frac{L}{2}|^{m_1}\leq U_\ell(x) \leq c_2|x+\frac{L}{2}|^{m_2} &\quad& \textrm{for  }x\leq
-\frac{L}{2} \\
c_1|x-\frac{L}{2}|^{m_1}\leq U_r(x) \leq c_2|x-\frac{L}{2}|^{m_2} &\quad&
\textrm{for  }x\geq
\frac{L}{2} \; 
\end{eqnarray}
for some constants $0<c_1<c_2$ and $2\leq m_1 < m_2<\infty$.
Recall that $U_\ell(x)=0$ for $x\geq -\frac{L}{2}$ and $U_r(x)=0$ for $x\leq
\frac{L}{2}$. We could allow steeper confinements but the present polynomial 
conditions turn out to be technically convenient.

We assume that each impurity is the source
of a local potential $V\in C^2$, \mbox{$0 \leq V(x,y) \leq V_0< \infty$}, $\supp V \subset
\mathbb{B}\Le(0,\frac{1}{4}\Ri)$, and that they are 
located at the sites of a finite lattice
 $\Lambda=\Le\{(n,m)\in \Z^2 ; n\in
[-\frac{L}{2}+\log L,\frac{L}{2}-\log L], m\in [-\frac{L}{2},\frac{L}{2}] 
\Ri\}$.
The random potential $V_\omega$ has the form
\begin{equation}\label{randompot}
V_\omega(x,y)=\sum_{(n,m)\in \Lambda} X_{n,m}(\omega) V(x-n,y-m)
\end{equation}
where the coupling constants
$X_{n,m}$ are i.i.d. random variables with common density 
$h \in C^2([-1,1])$  that satisfies
\mbox{$\|h\|_\infty < \infty$}, $\supp h=[-1, 1]$. 
We will denote by $\mathbb{P}_\Lambda$ the product measure defined on 
the set of all possible realizations 
\mbox{$\omega\in\Omega_{\Lambda}=[-1,1]^\Lambda$}. 
Clearly for any realization we have $|V_\omega(x,y)|\leq V_0$. Furthermore it
will be assumed that the random potential is weak in the sense that $4V_0<B$.

We will think of our system as being constituted of three 
pieces corresponding to
the \emph{bulk system} with the random Hamiltonian
\begin{equation}\label{2}
H_b=H_0+V_\omega
\end{equation}
and the \emph{left} and \emph{right edge systems} with non 
random Hamiltonians 
\begin{equation}\label{3}
H_\alpha=H_0 + U_\alpha, \qquad\qquad \alpha=\ell, r \; .
\end{equation}
All the Hamiltonians considered above have periodic boundary conditions
along the $y$ direction and are essentially self-adjoint 
on \mbox{$C_0^\infty(\R\times [-\frac{L}{2},\frac{L}{2}])$}.
For each realization
$\omega$ and size $L$ the spectrum $\sigma(H_\omega)$ of \eqref{h} (it depends on $L$)
consists of isolated eigenvalues of finite multiplicity.
In order to state our main result characterizing these eigenvalues we first have
to describe the spectra of \eqref{2} and \eqref{3}.

Let us begin with the edge Hamiltonians \eqref{3}. Here we state their properties without proofs
and refer the reader to \cite{MMP}, \cite{F} for more details.
Since the edge Hamiltonians $H_\alpha$ commute with $p_y$, 
they are decomposable into a direct sum
\begin{equation}
H_\alpha= \sideset{}{^\oplus}\sum_{k\in \frac{2\pi}{L}\Z} H_\alpha(k) = 
\sideset{}{^\oplus}\sum_{k\in \frac{2\pi}{L}\Z} \Le[\frac{1}{2}p_x^2 + \frac{1}{2}(k-Bx)^2 +
U_\alpha\Ri] \; .
\end{equation}
For each $k$ the one dimensional Hamiltonian $H_\alpha(k)$ has a 
compact resolvent, thus it has  discrete eigenvalues and by standard 
arguments one can show that they are not degenerate. If the $y$ direction would be infinitely 
extended, $k$ would vary over 
the real axis and the eigenvalues of $H_\alpha(k)$ would form spectral branches
$\varepsilon_\nu^\alpha(\hat{k})$, $\hat{k}\in \R$ labelled by the Landau level index $\nu$. These
 spectral branches are strictly monotone, entire functions with the properties
$\varepsilon_\nu^\ell(-\infty)=+\infty$, 
\mbox{$\varepsilon_\nu^\ell(+\infty)=(\nu
+\frac{1}{2})B$} and $ \varepsilon_\nu^r(-\infty)=(\nu +\frac{1}{2})B$,  
$\varepsilon_\nu^r(+\infty)=+\infty$. Here because of the 
periodic boundary conditions the set of $k$ values is discrete so that
the spectrum of $H_\alpha$
\begin{equation} 
\sigma(H_\alpha)=\Le\{E^\alpha_{\nu k}; \nu \in \N, k\in \frac{2\pi}{L}\Z\Ri\}
\end{equation}
consists of isolated points on the spectral branches 
$E^\alpha_{\nu k}= \varepsilon_\nu^\alpha(k)$, $k\in \frac{2\pi}{L}\Z$.
The corresponding eigenfunctions $\psi^\alpha_{\nu k}$ have the form
\begin{equation}
\psi^\alpha_{\nu k}(x,y)=\frac{1}{\sqrt{L}}e^{iky}\varphi^\alpha_{\nu k}(x)
\end{equation}
with $\varphi^\alpha_{\nu k}$ the normalized eigenfunctions of
the one-dimensional Hamiltonian $H_\alpha(k)$.
By definition, the current  of the state $\psi^\alpha_{\nu
k}$ in the $y$ direction
is given by the expectation value of the velocity $v_y=p_y-Bx$,
\begin{equation}\label{HF}
J^\alpha_{\nu k} = (\psi_{\nu k}^\alpha, v_y\psi_{\nu k}^\alpha)= 
\int_\R |\varphi^\alpha_{\nu k}(x)|^2(k-Bx) \D x 
=\partial_{\hat{k}} \varepsilon^\alpha_\nu(\hat{k})\Big|_{\hat{k}=\frac{2\pi m}{L}}
\end{equation}
where the last equality follows from the Feynman-Hellman theorem.
From \eqref{HF} we notice that for any $\varepsilon>0$, one can find
$j(\varepsilon)>0$ and $L(\varepsilon)$ such that for $L>L(\varepsilon)$
the states of the two branches $\nu=0$, $\alpha=\ell,r$ with energies $E_{0k}^\alpha\geq
\frac{1}{2}B+\varepsilon$ satisfy
\begin{equation}
J_{0k}^\ell\leq -j(\varepsilon)<0 \qquad \quad
J_{0k}^r\geq j(\varepsilon)>0 \; .
\end{equation}
In other words the eigenstates of the edge Hamiltonians carry an appreciable current.
The spacing of two consecutive
eigenvalues greater than $\frac{1}{2}B+\varepsilon$ satisfies
\begin{equation}
\Le|E^\alpha_{0\frac{2\pi(m+1)}{L}}-E^\alpha_{0\frac{2\pi
m}{L}}\Ri|>\frac{j(\varepsilon)}{L}\qquad
\alpha=\ell,r \; .
\end{equation}
Note that these observations extend to other branches but $j(\varepsilon)$ and $L(\varepsilon)$
 are  not 
uniform with respect to the index $\nu$. In the rest of the paper we limit ourselves to $\nu=0$
for simplicity.
On the other hand the spacing between the energies of $\sigma(H_\ell)$ and $\sigma(H_r)$ is a
priori arbitrary. We assume that the confining 
potentials $U_\ell$ and $U_r$ are
such that the following hypothesis is fulfilled.

\begin{hyp}\label{hyp1}
Fix any $\varepsilon>0$ and let
$\Delta_\varepsilon=\Le[\frac{1}{2}B+\varepsilon,\frac{1}{2}B+V_0\Ri]$. There exist
$L(\varepsilon)$ and $d(\varepsilon)>0$ such that for all $L>L(\varepsilon)$
\begin{equation}
\dist \Le(\sigma(H_\ell)\cap \Delta_\varepsilon,\sigma(H_r)\cap \Delta_\varepsilon\Ri)\geq
\frac{d(\varepsilon)}{L} \; .
\end{equation}
\end{hyp}

This hypothesis is important because a minimal amount of non-degeneracy 
between the spectra of the two edge systems is needed in order to
control backscattering effects induced by the random potential. Indeed in a system with 
two boundaries backscattering favors localization and has a tendency to destroy currents. 
This hypothesis can easily be realized by taking non-symmetric confining
potentials $U_\ell$ and $U_r$. In a more realistic model with impurities 
close to the edges one expects that it 
is automatically satisfied with a large probability.

Now we describe the spectral properties of the bulk random Hamiltonian \eqref{2}.
From the 
bound \eqref{est1} on the kernel of $R_0(z)$ and the fact that $V_\omega$
is bounded with compact support we can see that $V_\omega$ is relatively compact
w.r.t. $H_0$, thus 
\mbox{$\sigma_{ess}(H_b)=
\Le\{(\nu+\frac{1}{2})B; \nu\in \N\Ri\}$}. Since $|V_\omega(x,y)|\leq V_0<B$
the eigenvalues $E_\beta^b$ of 
$H_b$ are contained in Landau bands
$\bigcup_{\nu\geq 0} \Le[(\nu+\frac{1}{2})B - V_0,(\nu+\frac{1}{2})B +
V_0\Ri]$.
 We will assume 

\begin{hyp}\label{hyp2}
Fix any $\varepsilon > 0$. There exist $\mu(\varepsilon)$ a strictly positive constant and 
$L(\varepsilon)$ such that for all
$L>L(\varepsilon)$ one can find a set of realizations of the random potential
$\Omega_\Lambda^{'}$ with $\mathbb{P}_\Lambda(\Omega_\Lambda^{'}) \geq 1- L^{-\theta}$,
$\theta>0$,
with the property that if
$\omega\in \Omega_\Lambda^{'}$ the 
eigenstates corresponding to $E_\beta^b \in \sigma(H_b)\cap \Delta_\varepsilon$  satisfy
\begin{equation}\label{H2}
|\psi_\beta^b(x,\bar{y}_\beta)|\leq e^{-\mu(\varepsilon) L} \qquad,\qquad
|\partial_y\psi_\beta^b(x,\bar{y}_\beta)|\leq e^{-\mu(\varepsilon) L} 
\end{equation}
for some $\bar{y}_\beta$ depending on $\omega$ and $L$.
\end{hyp}

Since $V_\omega$ is random we expect that wavefunctions with energies in $\Delta_\varepsilon$
(not too close to the Landau levels where the localization length diverges) are exponentially
localized on a scale ${\cal O}(1)$ with respect to $L$. Inequalities \eqref{H2} are a weaker version 
of this statement, and have been checked for the special case where the random potential 
is a sum of rank one perturbations \cite{FM2} using the methods of Aizenman and
Molchanov \cite{AM} (see for example \cite{DMP4} where the case of point
impurities is treated by these methods).
Presumably
one could adapt existing techniques for multiplicative potentials to our geometry, to prove hypothesis $(H2)$ at least 
for energies
close to the band tail $\frac{B}{2} +V_0$. One also expects that
$\mu(\varepsilon)\to 0$ as $\varepsilon\to 0$.
The main physical consequence of $(H2)$ (as shown in section 5) is that  
a state satisfying \eqref{H2} does not carry any appreciable current 
(contrary to the eigenstates of $H_\alpha$) in the sense that
\mbox{$J_\beta^b=(\psi_\beta^b, v_y\psi_\beta^b)={\cal O}(e^{-\mu(\varepsilon) L})$}.

We now state our main result.

\begin{thm}
Fix $\varepsilon>0$ and assume that $(H1)$ and $(H2)$ are fulfilled. Assume $B>4V_0$. 
Let $p\geq 7$ and $s=\min(\theta,p-6)$.  
Then there exists a numerical constant $\gamma>0$ and an $L(\varepsilon,p,B,V_0)$ such that for all
for all $L>L(\varepsilon,p,B,V_0)$ one can find a set $\hat{\Omega}_\Lambda$ of 
realizations of the
random potential with $\mathbb{P}_\Lambda(\hat{\Omega}_\Lambda)\geq 1- 3L^{-s}$
 such that for any
$\omega\in \hat{\Omega}_\Lambda$, $\sigma(H_\omega)\cap \Delta_\varepsilon$ is the union of
three sets $\Sigma_{\ell}\cup \Sigma_{b}\cup \Sigma_{r}$, each depending
on $\omega$ and $L$, and characterized by the following properties:
\begin{enumerate}
\item[a)] $E^\alpha_k\in\Sigma_\alpha$ $(\alpha=\ell,r)$ are a small
perturbation of $E^\alpha_{0k}\in\sigma(H_\alpha)\cap \Delta_\varepsilon$ with 
\begin{equation}\label{rt1}
|{E}^\alpha_{k}-E^\alpha_{0k}|\leq  e^{-\gamma B(\log L)^2}, \qquad \qquad
\alpha=\ell, r \; .
\end{equation}
\item[b)] For ${E}^\alpha_{k}\in \Sigma_\alpha$ the current $J^\alpha_{k}$ of 
the associated eigenstate
satisfies 
\begin{equation}\label{rt2}
\Le|J^\alpha_{k}-J^\alpha_{0k}\Ri|\leq e^{-\gamma B(\log L)^2}, \qquad \qquad
\alpha=\ell, r \; .
\end{equation}
\item[c)] $\Sigma_b$ contains the same number of energy levels as
$\sigma(H_b)\cap \Delta_\varepsilon$ and 
\begin{equation}\label{rt3}
\dist(\Sigma_b,\Sigma_\alpha) \geq L^{-p+1}, \qquad \qquad
\alpha=\ell, r \; .
\end{equation}
\item[d)] The current associated to each level $E_\beta\in \Sigma_b$ satisfies
\begin{equation}\label{rt4}
|J_\beta|\leq  e^{-\gamma B(\log L)^2}\; .
\end{equation}
\end{enumerate}
\end{thm}

The proof of the theorem is organized as follows. In section 3 we set up a 
decoupling scheme by which we express the resolvent of $H_\omega$ as an 
approximate sum of those of the edge and bulk systems. Parts $a)$ and $c)$
of Theorem 1 are proven in section 4.
First we compute 
approximations for the spectral projections of $H_\omega$ in terms of the 
projectors $P(E^\alpha_{0k})$ of $H_\alpha$ and $P_b(\bar{\Delta})$ of
$H_b$ (Proposition 1). This is done for realizations of the disorder such
that the levels of $H_b$ are not ``too close'' to those of $H_\alpha$. We then 
show that these realizations are typical (have large probability) thanks to
a Wegner estimate (Proposition 2). Parts $b)$ and $d)$ are proven in section 5 
by estimating currents in term of norms of differences between projectors. The appendices
contain some technical estimates.

\section{Decoupling of the Bulk and the Edge Systems}

The resolvent $R(z)=(z-H_\omega)^{-1}$ can be expressed, up to a
small term, as a sum of the resolvents of the bulk system
$R_b(z)=(z-H_b)^{-1}$ and the two edge systems
$R_\alpha(z)=(z-H_\alpha)^{-1}$ ($\alpha=\ell,r$). Here this will be achieved 
by a {\it decoupling formula}  
developed in other contexts  \cite{BCD}, \cite{BG}. 
We set $D =\log L$ and introduce the characteristic functions
\begin{eqnarray}
\tilde{J}_\ell(x)&=&\chi_{]-\infty,\mbox{\tiny$-\frac{L}{2}+\frac{D}{2}$}]}(x) \qquad 
\tilde{J}_b(x)=\chi_{[\mbox{\tiny$-\frac{L}{2}+\frac{D}{2}$},
\mbox{\tiny$\frac{L}{2}-\frac{D}{2}$}]}(x)
\nonumber\\ 
\tilde{J}_r(x)&=&\chi_{[\mbox{\tiny$\frac{L}{2}-\frac{D}{2}$},+\infty[}(x) \;
.
\end{eqnarray}
We will also use three bounded $C^\infty(\R)$ functions $|J_i(x)|\leq 1$, $i\in {\cal
I}\equiv \{\ell,b,r\}$, with bounded first and second
derivatives $\sup_{x}|\partial_x^nJ_i(x)|\leq 2$, $n=1,2$,  
and such that
\begin{eqnarray}
& &J_\ell(x)=
\begin{cases}
1 \quad \text{if } x\leq -\frac{L}{2}+\frac{3D}{4} \\
0 \quad \text{if } x\geq -\frac{L}{2}+\frac{3D}{4}+1\\
\end{cases} \qquad
J_b(x)=
\begin{cases}
1 \quad \text{if }  |x|\leq \frac{L}{2}-\frac{D}{4} \\
0 \quad \text{if }  |x|\geq \frac{L}{2}-\frac{D}{4}+1 \\
\end{cases} 
\nonumber\\ 
& &J_r(x)=
\begin{cases}
1 \quad \text{if } x\geq \frac{L}{2}-\frac{3D}{4} \\
0 \quad \text{if } x\leq \frac{L}{2}-\frac{3D}{4}-1 \\
\end{cases} \; .
\end{eqnarray}
For $i\in {\cal I}$ we have $H_\omega J_i=H_iJ_i$ thus
\begin{equation}\label{bg}
(z-H_\omega)\sum_{i\in {\cal I}} J_i R_i(z) \tilde{J}_i
=\sum_{i\in {\cal I}}(z-H_i)J_i R_i(z) \tilde{J}_i
=1-{\cal K}(z)
\end{equation}
where 
\begin{equation}
{\cal K}(z)=\sum_{i\in {\cal I}}K_i(z)=\sum_{i\in {\cal I}}
\frac{1}{2}[p_x^2,J_i]R_i(z)\tilde{J}_i \; .
\end{equation}
To obtain the second equality one commutes $(z-H_i)$ and $J_i$ and then uses the identity
$\sum_{i\in {\cal I}}J_i\tilde{J}_i=\sum_{i\in {\cal I}}\tilde{J}_i=1$.
From \eqref{bg} we deduce the decoupling formula
\begin{equation}\label{resolvent}
R(z)=\Le(\sum_{i\in {\cal I}} J_i R_i(z) \tilde{J}_i\Ri)\Le( 1 - 
{\cal K}(z)\Ri)^{-1} \; .
\end{equation}

The main result of this section is an estimate of the operator norm of 
${\cal K}(z)$. In particular it will assure
$\|{\cal K}(z)\|<1$.

\begin{lem}\label{prop8}
Let ${\cal R}e\, z\in \Delta_\varepsilon$ such that
$\dist(z,\sigma(H_\ell)\cup\sigma(H_r)\cup\sigma(H_b))
\geq e^{-\frac{B}{512}(\log L)^2}$. 
Then for $L$ large
enough there exists a constant $C(B,V_0)>0$ independent of $L$ such that
\begin{equation}
\|{\cal K}(z)\| \leq \varepsilon^{-1}C(B,V_0)L e^{-\frac{B}{512}(\log L)^2}\; .
\end{equation}
\end{lem}

\begin{proof}
Computing the commutator in the definition of $K_i(z)$ and applying the 
second resolvent formula we have 
\begin{eqnarray}
K_i(z)&=&- \frac{1}{2}(\partial_x^2J_i)R_i(z)\tilde{J}_i -
(\partial_xJ_i)\partial_xR_i(z)\tilde{J}_i\nonumber \\
&=&
-\frac{1}{2}(\partial_x^2J_i)R_0(z)\tilde{J}_i -
\frac{1}{2}(\partial_x^2J_i)R_0(z)W_iR_i(z)\tilde{J}_i \nonumber \\
&-& (\partial_xJ_i)\partial_xR_0(z)\tilde{J}_i 
- (\partial_xJ_i)\partial_xR_0(z)W_iR_i(z)\tilde{J}_i
\end{eqnarray}
where we have set $W_\ell=U_\ell$, $W_b=V_\omega$ and $W_r=U_r$.
From the triangle inequality and $\|R_i(z)\|=\dist(z,\sigma(H_i))^{-1}$
we obtain
\begin{eqnarray}\label{c}
\| K_i(z)\| &\leq&  \frac{1}{2}\|(\partial^2_xJ_i)R_0(z)\tilde{J}_i\| +
\frac{1}{2}\|(\partial^2_xJ_i)R_0(z)W_i\|\,\dist(z,\sigma(H_i))^{-1} \nonumber \\
&+& \|(\partial_xJ_i)\partial_xR_0(z)\tilde{J}_i\| 
+ \|(\partial_xJ_i)\partial_xR_0(z)W_i\|\,\dist(z,\sigma(H_i))^{-1} \; .
\end{eqnarray}
To estimate the operator norms on the right hand side it is sufficient to bound them
by the Hilbert-Schmidt norms $\|.\|_{2}$. Using bounds \eqref{est1} on the kernels of
$\partial_x^nR_0(z)$ for $n=0,1$, and the properties of the functions $J_i$, $\tilde J_i$ we obtain
\begin{eqnarray}
\|(\partial_x^{2-n}J_i) \partial_x^n 
R_0(z)\tilde{J}_i\|^2_{2}  
&=& \int_{\supp \partial_x^{2-n}J_i} \textrm{d} \bold{x} 
|\partial_x^{2-n}J_i(x)|^2 \int_{\supp \tilde{J}_i} \textrm{d} \bold{x}'
|\partial_x^n R_0(\bold{x},\bold{x}';z)|^2
\nonumber\\
&\leq& 4C_n^2(z,B)\int_{\supp \partial_x^{2-n}J_i} \textrm{d} \bold{x} 
\int_{\supp \tilde{J}_i} \textrm{d} \bold{x}' e^{-\frac{B}{4}(x-x')^2} 
\nonumber\\ &\leq& 4C_n^2(z,B)e^{-\frac{B}{8}\Le(\frac{D}{4}+1\Ri)^2}
\int_{\supp \partial_x^{2-n}J_i} \textrm{d} \bold{x} 
\int_{\R\times [-\frac{L}{2},\frac{L}{2}]} \textrm{d} \bold{x}'e^{-\frac{B}{8}(x-x')^2} \nonumber \\
&\leq& 16\sqrt{\frac{\pi}{B}}C_n^2(z,B)L^2e^{-\frac{B}{128}D^2} \; . \label{b}
\end{eqnarray}
For the norms involving the potentials $W_i$ we obtain in a similar way
\begin{eqnarray}\label{310}
& &\|\partial_x^{2-n}J_i \partial_x^n  R_0(z)W_i\|^2_{2} \nonumber \\
&=& \int_{\supp \partial_x^{2-n}J_i} \textrm{d} \bold{x}|\partial_x^{2-n}J_i(x)|^2 
\int_{\supp W_i} \textrm{d} \bold{x}' |\partial_x^\alpha
R_0(\bold{x},\bold{x}';z)|^2
|W_i(\bold{x}')|^2 \nonumber\\
&\leq& 4C^2_n(z,B)e^{-\frac{B}{128}D^2}\int_{\supp \partial_x^{2-n}J_i} \textrm{d} \bold{x} 
 \int_{\supp W_i} \textrm{d}
\bold{x}'e^{-\frac{B}{8}(x-x')^2}|W_i(\bold{x}')|^2 \; .
\end{eqnarray} 
It is clear that since $V_\omega$ is bounded, and $U_\ell$, $U_r$ do not grow 
faster than polynomials, the double integral in the right hand side of the last
inequality is bounded above by $L^2$ times a constant depending only on $B$
and $V_0$. From  this result, \eqref{c}, \eqref{b} and 
 $\dist(z,\sigma(H_\ell)\cup\sigma(H_r)\cup\sigma(H_b))
\geq e^{-\frac{B}{512}(\log L)^2}$ we obtain ($\tilde{C}(B,V_0)$ a constant independent of $L$)
\begin{equation}
\|K_i(z)\|\leq \tilde{C}(B, V_0)\varepsilon^{-1}Le^{-\frac{B}{512}(\log L)^2} \;
,
\end{equation}
where we used the expression for $C_n(z,B)$
in Appendix A and the fact that ${\cal R}e\, z\in \Delta_\varepsilon$.

\end{proof}

\section{Estimates of Eigenprojectors of $H_\omega$}

In this section we use the decoupling formula \eqref{resolvent} to give
deterministic estimates for the difference between projectors of $H_\omega$ and
 $H_b$, $H_\ell$ and $H_r$. We then
combine this information with a probabilistic estimate 
(Wegner estimate) to
deduce that the spectrum of $H_\omega$
is the union of the three sets $\Sigma_\ell$,
$\Sigma_r$ and $\Sigma_b$ satisfying the
parts $a)$ and $c)$ of Theorem 1. 

\begin{prop}\label{thm1}
Assume that $(H1)$ holds. Take \mbox{$p\geq 7$} and any $e^{-\frac{B}{512}(\log
L)^2}<\rho<
\frac{d(\varepsilon)}{2}L^{-p}$. 
For $L>L(\varepsilon)$ let $\Omega^{''}_\Lambda$ be the set of
realizations of the random potential such that for each $\omega\in
\Omega_\Lambda''$
$\dist\Le(\sigma(H_b)\cap \Delta_\varepsilon,E_{0k}^\alpha \Ri)\geq
d(\varepsilon)L^{-p}$ for all $E_{0k}^\alpha\in \Delta_\varepsilon$,
$\alpha=\ell,r$. Then

\begin{enumerate}
\item[i)] If $P(E^\alpha_{0k})$  is the eigenprojector of $H_\alpha$ associated to
the eigenvalue \mbox{$E^\alpha_{0k}\in \Delta_\varepsilon$} and $P_k^\alpha$ the 
eigenprojector of $H_\omega$ for the intervals 
$I_k^\alpha=[E^\alpha_{0k}-\rho, E^\alpha_{0k}+\rho]$ 
 we have
\begin{equation}\label{t1}
\|P_k^\alpha-P(E^\alpha_{0k})\| \leq
\varepsilon^{-1}C'(B,V_0)Le^{-\frac{B}{512}(\log L)^2} \; .
\end{equation}
\item[ii)] Let $\bar{\Delta}\subset \Delta_\varepsilon$ be an interval such that
$\dist(\bar{\Delta},\sigma(H_\ell)\cup\sigma(H_r))= \frac{d(\varepsilon)}{2}L^{-p}$.
If $P_b(\bar{\Delta})$ is the eigenprojector of $H_b$ for the interval
$\bar{\Delta}$ and $P(\bar{\Delta})$ the eigenprojector
of $H_\omega$ for the interval $\bar{\Delta}$ we have
\begin{equation}\label{t3}
\|P(\bar{\Delta})-P_b(\bar{\Delta})\| \leq \varepsilon^{-3}C'(B,V_0)L^{p}
e^{-\frac{B}{512}(\log L)^2} \; .
\end{equation}
\end{enumerate}
\end{prop}

\begin{proof}
We start by proving \eqref{t1} for $\alpha=r$. The case $\alpha=\ell$ is identical. 
From the decoupling formula we have
\begin{eqnarray}\label{dec}
R(z)-R_r(z) &=&\Le(\sum_{i\in {\cal I}} J_iR_i(z)\tilde{J}_i\Ri)
\Le(\sum_{n=1}^\infty {\cal K}(z)^n\Ri) -(1-J_r)R_r(z) \nonumber\\
&-& J_rR_r(z)(1-\tilde{J}_r)
+ J_\ell R_\ell(z) \tilde{J}_\ell + J_bR_b(z)\tilde{J}_b \; .
\end{eqnarray}
Let $\Gamma$ be a circle of radius $\rho$ in the complex plane, centered at $E_{0k}^r$.
Because of $(H1)$ and $\dist\Le(\sigma(H_b)\cap \Delta_\varepsilon,
E^r_{0k}\Ri)\geq d(\varepsilon)L^{-p}$, $R_b(z)$ and $R_\ell(z)$ have no poles
in $\Gamma$.
Moreover the only pole of $R_r(z)$ is precisely $E_{0k}^r$. Thus integrating
\eqref{dec} along the circle $\Gamma$
\begin{eqnarray}\label{dpr}
P_k^r-P(E^r_{0k}) &=& \frac{1}{2\pi i}\oint_{\Gamma} \Le(\sum_{i\in {\cal
I}} J_iR_i(z)\tilde{J}_i\Ri)
\sum_{n=1}^\infty {\cal K}(z)^n \D z\nonumber\\
&-&(1-J_r)P(E^r_{0k})-J_rP(E^r_{0k})(1-\tilde{J}_r) \; .
\end{eqnarray}
We proceed to estimate the norms of the three contributions on the right hand 
side of \eqref{dpr}. The norm of the first term is smaller than
\begin{eqnarray}\label{a}
\rho\Le(\sum_{i\in {\cal I}}\sup_{z\in \Gamma}\|R_i(z)\|\Ri)
\frac{\sup_{z\in \Gamma} \|{\cal K}(z)\|}{1- \sup_{z\in \Gamma} \|{\cal K}(z)\|}
\leq 6\varepsilon^{-1}C(B,V_0)Le^{-\frac{B}{512}(\log L)^2} \; .
\end{eqnarray}
Indeed, for $i=r$ we have $\sup_{z\in \Gamma}\|R_r(z)\|=\rho^{-1}$ by construction. For $i=\ell,b$ we have
$\sup_{z\in \Gamma}\|R_i(z)\|<\frac{2}{d(\varepsilon)}L^{p}$. Since $\rho<
\frac{d(\varepsilon)}{2}L^{-p}$ we note that in all three cases ($i\in {\cal I}$)
$\rho \sup_{z\in \Gamma}\|R_i(z)\|\leq 1$. Furthermore, since $\rho> e^{-\frac{B}{512}(\log L)^2}$,
using Lemma 1 we get \eqref{a}.
To estimate the second term in \eqref{dpr} we note that by the second resolvent formula
\begin{equation}\label{for}
\frac{P(E^r_{0k})}{(z-E^r_{0k})}=(z-H_0)^{-1}P_r(E^r_{0k}) + 
(z-H_0)^{-1}U_r\frac{P(E^r_{0k})}{(z-E^r_{0k})} \; .
\end{equation}
Integrating \eqref{for} along $\Gamma$ we obtain the identity
\begin{equation}\label{le}
P(E^r_{0k})=(E^r_{0k}-H_0)^{-1}U_rP(E^r_{0k})
\end{equation}
this implies
\begin{eqnarray}
\|(1-J_r)P(E^r_{0k})\| &\leq& \|(1-J_r)R_0(E^r_{0k})U_r\| 
\leq\|(1-J_r)R_0(E^r_{0k})U_r\|_{2} \\
&=& \Le\{\int \D \bold{x} |1-J_r(x)|^2 \int \D
\bold{x}'|R_0(\bold{x},\bold{x}';E^r_{0k})U_r(x')|^2 \Ri\}^{1/2}\nonumber
\end{eqnarray}
since the distance (in the $x$ direction) between the supports of $(1-J_r)$ and $U_r$ 
is greater than
$\frac{D}{2}+1$ we can proceed in a similar way as in the estimate of
\eqref{310} to obtain
\begin{equation}\label{bbb}
\|(1-J_r)P(E^r_{0k})\| \leq \varepsilon^{-1}\bar{C}(B)Le^{-\frac{B}{64}(\log L)^2}
\end{equation}
where $\bar{C}(B)$ is a constant depending only on $B$.
For the third term in \eqref{dpr} we use the adjoint of \eqref{le}
\begin{equation}
P(E^r_{0k})=P(E^r_{0k})U_r(E^r_{0k}-H_0)^{-1}
\end{equation}
to get
\begin{eqnarray}
\|J_rP(E^r_{0k})(1-\tilde{J}_r)\| \leq \|U_rR_0(E^r_{0k})(1-\tilde{J}_r)\|
\end{eqnarray}
from which we obtain the same bound as in \eqref{bbb}. Combining this result with \eqref{dpr},
\eqref{a}, \eqref{bbb} we obtain \eqref{t1} in the proposition.

Let us now sketch the proof of \eqref{t3}. From the 
decoupling formula we have
\begin{eqnarray}
R(z)-R_b(z) &=&\Le(\sum_{i\in {\cal I}} J_iR_i(z)\tilde{J}_i\Ri)
\Le(\sum_{n=1}^\infty {\cal K}(z)^n\Ri) 
-(1-J_b)R_b(z)\nonumber\\ &-& J_bR_b(z)(1-\tilde{J}_b) 
+ J_\ell R_\ell(z)\tilde{J}_\ell + J_rR_r(z)\tilde{J}_r\; .
\end{eqnarray}
Given an interval $\bar{\Delta}\subset \Delta_\varepsilon$ such that
$\dist(\bar{\Delta},\sigma(H_\ell)\cup\sigma(H_r))= \frac{d(\varepsilon)}{2}L^{-p}$, 
we choose a circle $\bar{\Gamma}$ in the complex plane with diameter equal to
$|\bar{\Delta}|$. Then if we integrate over $\bar{\Gamma}$ the last two terms on the 
right hand side do not contribute while the second and third ones give
$(1-J_b)P_b(\bar{\Delta})$ and $J_bP_b(\bar{\Delta})(1-J_b)$. Therefore
\begin{eqnarray}\label{pdb}
\|P-P_b(\bar{\Delta})\| &\leq& |\bar{\Delta}|\Le(\sum_{i\in {\cal I}}\sup_{z\in
\bar{\Gamma}}\|R_i(z)\|\Ri)
\frac{\sup_{z\in \bar{\Gamma}} \|{\cal K}(z)\|}{1- \sup_{z\in \bar{\Gamma}} \|{\cal K}(z)\|} \nonumber\\
&+& \|(1-J_b)P_b(\bar{\Delta})\|+\|J_bP_b(\bar{\Delta})(1-\tilde{J}_b)\| \; .
\end{eqnarray}
From Lemma 1, $|\bar{\Delta}|<d(\varepsilon)L^{-1}$ and 
$\sup_{z\in \bar{\Gamma}}\|R_i(z)\|<\frac{2}{d(\varepsilon)}L^{p}$
the first term is bounded above by 
\begin{equation}\label{fb}
12\varepsilon^{-1}C(B,V_0)L^{p}e^{-\frac{B}{512}(\log L)^2} \; .
\end{equation}
In order to
estimate the second  norm in \eqref{pdb}
we notice that (in the same way as in \eqref{for}, \eqref{le})
\begin{equation}\label{id}
P_b(\bar{\Delta})=\sum_{E_\beta^b\in \bar{\Delta}}
R_0(E_\beta^b)V_\omega P_b(E_\beta^b)
\end{equation}
thus
\begin{equation}
\|(1-J_b)P_b(\bar{\Delta})\|\leq \sum_{E_\beta^b\in \bar{\Delta}}
\|(1-J_b)R_0(E_\beta^b)V_\omega\|_{2} \; .
\end{equation}
Each term of the sum can be bounded in a way similar to \eqref{310},
and since the number of terms in the sum is equal to $\Tr P_b(\bar{\Delta})$ we get
\begin{eqnarray}\label{bwn}
\|(1-J_b)P_b(\bar{\Delta})\| &\leq& \varepsilon^{-1}C(B,V_0)Le^{-\frac{B}{64}(\log L)^2}\Tr P_b(\bar{\Delta})
\nonumber\\
&\leq& 2\varepsilon^{-3}c(B)^2C(B,V_0)V_0^2L^5e^{-\frac{B}{64}(\log L)^2} \; .
\end{eqnarray}
The second inequality follows from Lemma 4 in Appendix B (where we need $B>4V_0$).
For $\|J_bP_b(\bar{\Delta})(1-\tilde{J}_b)\|$ one uses the adjoint of identity \eqref{id} 
to obtain the same result. The result \eqref{t3} of  the proposition then follows by
combining \eqref{pdb}, \eqref{fb} and \eqref{bwn}.
\end{proof}

In appendix B we adapt the method of \cite{CH} to our geometry to get the 
following Wegner estimate.

\begin{prop}\label{prop3}
Let $B\geq 4V_0$ and $E\in \Delta_\varepsilon$
\begin{equation}
\mathbb{P}_\Lambda\Le(\dist(\sigma(H_b),E)<\delta\Ri) \leq 
4c(B)\|h\|_{\infty}\delta\varepsilon^{-2} V_0 L^4 \; .
\end{equation}
\end{prop}

\begin{proof}[Proof of Theorem 1, part a) and c)] 
Let $\omega \in \Omega_\Lambda^{''}$ where $\Omega_\Lambda^{''}$ is the set given in Proposition
\ref{thm1}. Since for $L$ large enough the right hand side of \eqref{t1} is strictly smaller
than one
the two projectors  necessarily 
have the same dimension. Therefore $\sigma(H_\omega)\cap
I^\alpha_k$ contains a unique energy level $E_k^\alpha$ for each
$I^\alpha_k$ of radius $\rho$. In particular by taking the smallest 
value $\rho=e^{-\frac{B}{512}(\log L)^2}$ we get \eqref{rt1}. 
The number of such levels is ${\cal
O}(L)$ since they are in one to one correspondence with the energy levels of
$H_\alpha$. The sets $\Sigma_\alpha$ of Theorem 1 are precisely
\begin{equation}
\Sigma_\alpha = \bigcup_{k} \Le(\sigma(H_\omega)\cap I^\alpha_k\cap
\Delta_\varepsilon\Ri),\qquad\qquad \alpha=\ell,r \; .
\end{equation}
The set of all other eigenvalues in $\sigma(H_\omega)\cap
\Delta_\varepsilon$, defines $\Sigma_b$, and is necessarily
contained in intervals $\bar{\Delta}$ such that
$\dist(\bar{\Delta}, \sigma(H_\ell)\cup\sigma(H_r))=\frac{d(\varepsilon)}{2}L^{-p}$. 
In view of \eqref{rt1} this implies \eqref{rt3}.
Since the two projectors in \eqref{t3} necessarily have the same dimension, the number of
eigenstates in $\Sigma_b$ is the same than that of $\sigma(H_b)\cap\Delta_\varepsilon$.
It remains to estimate the probability of the set $\Omega_\Lambda^{''}$.
The realizations of the complementary set are such that 
for at least one $E_{0k}^\alpha\in \Delta_\varepsilon$
\begin{equation}
\dist(\sigma(H_b),E_{0k}^\alpha)<d(\varepsilon)L^{-p}
\end{equation}
but from Proposition \ref{prop3} this has a probability smaller than 
\begin{equation}
4c(B)\|h\|_{\infty}d(\varepsilon)L^{-p}\varepsilon^{-2}V_0 L^4 \cdot
{\cal O}(L)
\end{equation}
where ${\cal O}(L)$ comes from the number of levels in
$[\sigma(H_\ell)\cup\sigma(H_r)]\cap \Delta_\varepsilon$. Thus for $L$ large enough
\begin{equation}
\mathbb{P}_\Lambda(\Omega_\Lambda^{''})\geq 1 -  L^{6-p} \; .
\end{equation}
We recall that $p\geq 7$.
\end{proof}

\section{Estimates of Currents}

In this section we characterize the eigenvalues of $H_\omega$ in terms of
the current carried by the  corresponding eigenstates. This
will yield parts $b)$ and $d)$ of Theorem 1.

\begin{proof}[Proof of Theorem 1, part b)] 
Let ${E}^\alpha_k\in \Sigma_\alpha$. The  associated current is
by definition 
\begin{equation}
J^\alpha_k=\Tr v_yP^\alpha_k
\end{equation}
and will be compared to that of $\psi_{0k}^\alpha$
\begin{equation}
J^\alpha_{0k}=\Tr v_yP(E^\alpha_{0k}) \; .
\end{equation}
The difference between these two currents will be 
estimated by $\|P^\alpha_k-P(E^\alpha_{0k})\|$. First we observe that
$v_yP^\alpha_k$ is trace class. Indeed,
$v_yP^\alpha_k=v_yP^\alpha_kP^\alpha_k$ with $v_yP^\alpha_k$ bounded and 
$\|P^\alpha_k\|_1=\Tr P^\alpha_k =1$
\begin{equation}\label{tn1}
\|v_yP^\alpha_k\|_1^2\leq\|v_yP^\alpha_k\|^2\leq \|P^\alpha_k v_y^2 P^\alpha_k\|\leq 
2\|P^\alpha_k (H_\omega-V_\omega) P^\alpha_k\|\leq 2E_k^\alpha+V_0
\end{equation}
to get the second inequality one has simply added positive terms to 
$v_y^2$.
Similarly
\begin{eqnarray}\label{tn2}
\|v_yP(E^\alpha_{0k})\|_1^2&\leq&\|v_yP(E^\alpha_{0k})\|^2\leq \|P(E^\alpha_{0k}) v_y^2 P(E^\alpha_{0k})\|
\nonumber \\
&\leq& 
2\|P(E^\alpha_{0k}) H_\alpha P(E^\alpha_{0k})\|\leq 2E_{0k}^\alpha\; .
\end{eqnarray}
The identity
\begin{eqnarray}\label{iden}
P_k^\alpha-P(E_{0k}^\alpha) &=& 
[P_k^\alpha-P(E_{0k}^\alpha)]^2 + 
[P_k^\alpha-P(E_{0k}^\alpha)]P(E_{0k}^\alpha) \nonumber \\
&+& P(E_{0k}^\alpha)[P_k^\alpha-P(E_{0k}^\alpha)]
\end{eqnarray}
implies
\begin{eqnarray}\label{equal}
|J^\alpha_k-J^\alpha_{0k}|&=&
\Le|\Tr v_y[P_\alpha^k-P(E_{0k}^\alpha)]\Ri|
\leq \Le|\Tr v_y[P_k^\alpha-P(E_{0k}^\alpha)]^2\Ri| \nonumber\\ 
&+&\Le|\Tr v_y[P_k^\alpha-P(E_{0k}^\alpha)]P(E_{0k}^\alpha)\Ri|\nonumber \\
&+&\Le|\Tr v_yP(E_{0k}^\alpha)[P_k^\alpha-P(E_{0k}^\alpha)]\Ri| \; .
\end{eqnarray}
From \eqref{equal}, \eqref{tn1} and \eqref{tn2} we get
\begin{eqnarray}
|J^\alpha_k-J^\alpha_{0k}| &\leq& 2 \Le( 
\|v_yP_k^\alpha\|_1 + \|v_yP(E_{0k}^\alpha)\|_1 \Ri)
\|P_k^\alpha-P(E_{0k}^\alpha)\| \nonumber \\ 
&\leq&
2\Le((B+3V_0)^{1/2}+(B+2V_0)^{1/2}\Ri) \|P_k^\alpha-P(E_{0k}^\alpha)\| \; .
\end{eqnarray}
Combining this last inequality with \eqref{t1} we get the result
\eqref{rt2} of Theorem 1.
\end{proof}

In order to prove part $d)$ of Theorem 1 we need the following lemma.

\begin{lem}\label{LC}
Fix $\omega\in\Omega_\Lambda^{'}$ the set of realizations in $(H2)$.
Let $\psi^b_1$, $\psi^b_2$ be two eigenstates of $H_b$ with eigenvalues $E_1^b$
and $E^b_2$. Then
\begin{equation}\label{fff}
|(\psi^b_1,v_y\psi^b_2)| \leq 2|E^b_1-E^b_2|L +  e^{-\frac{\mu(\varepsilon)}{4}
L}\; .
\end{equation} 
\end{lem}

For $\psi^b_1=\psi^b_2$, $E^b_1=E^b_2$ this shows that eigenstates of 
$H_b$ do not carry any appreciable current. The main idea of the proof
sketched below is that $v_y$ is equal to the commutator $[-iy,H_b]$ up to a small
boundary term.

\begin{proof}
The wavefunctions $\psi_1^b$ and $\psi_2^b$ are defined on 
$\R\times [-\frac{L}{2},\frac{L}{2}]$, are periodic along $y$ and are twice differentiable in $y$.
Here we will work with periodized versions of these functions
 where the $y$ direction is infinite
(but we keep the same notation). This allows us to shift  integrals
over $y$  from $[-\frac{L}{2},\frac{L}{2}]$ to $[\bar{y}_2,\bar{y}_2+L]$.
We have
\begin{equation}
(\psi^b_1, v_y\psi^b_2)=\int_\R \textrm{d}x 
\int_{\bar{y}_2}^{\bar{y}_2+L} \textrm{d}y  [\psi^b_1(\bold{x})]^*
(-i\partial_y-Bx)\psi^b_2(\bold{x}) \; .
\end{equation}
An integration by parts yields
\begin{eqnarray}\label{per}
i(\psi^b_1,v_y\psi^b_2) &=&\frac{1}{2}\int_\R \textrm{d}x 
\int_{\bar{y}_2}^{\bar{y}_2+L} \textrm{d}y
[\psi^b_1(\bold{x})]^{*}y(-i\partial_y-Bx)^2\psi^b_2(\bold{x})\nonumber\\
&-&\frac{1}{2}\int_\R \textrm{d}x
\int_{\bar{y}_2}^{\bar{y}_2+L} \textrm{d}y 
[(-i\partial_y-Bx)^2\psi^b_1(\bold{x})]^* y \psi^b_2(\bold{x}) 
+ {\cal B} 
\end{eqnarray}
where $\cal B$ is a boundary term given by
\begin{eqnarray}
{\cal B}&=&i\frac{L}{2}\int_\R \textrm{d}x [(-i\partial_y-Bx)\psi^b_1(x,\bar{y}_2)]^*
\psi^b_2(x,\bar{y}_2) \nonumber \\
&+&[\psi^b_1(x,\bar{y}_2)]^*(-i\partial_y-Bx)\psi^b_2(x,\bar{y}_2) \; .
\end{eqnarray}
We can add a periodized version of $V_\omega$ and $\frac{1}{2}p_x^2$ to the kinetic energy operator in
both terms on the right hand side of \eqref{per} and use that $\psi_1^b$ and
$\psi^b_2$ are eigenfunctions of $H_b$ to obtain
\begin{eqnarray}
i(\psi^b_1,v_y\psi^b_2) = (E^b_2-E^b_1) \int_\R \textrm{d}x 
\int_{\bar{y}_2}^{\bar{y}_2+L} \textrm{d}y  y [\psi^b_1(\bold{x})]^* \psi^b_2(\bold{x})
 + {\cal B}\; .
\end{eqnarray}
From $|y|\leq |\bar{y}_2|+L\leq 2L$ and the Schwarz inequality we obtain
\begin{equation}
|(\psi_1^b,v_y\psi_2^b)|\leq 2L |E_2^b-E_1^b|+|{\cal B}|\; .
\end{equation}
With the help of \eqref{LL1}, \eqref{LL2} in Appendix C we get
\begin{equation}
|{\cal B}|\leq e^{-\frac{\mu(\varepsilon)}{4}L}
\end{equation}
this concludes the proof of \eqref{fff}. 
\end{proof}

\begin{proof}[Proof of Theorem 1, part d)]
Let $\bar{\Delta}$ an interval like in part $ii)$ of Proposition \ref{thm1}.
We consider the maximal set of
intervals ${\cal F}_k\subset \bar{\Delta}$ such that 
$|{\cal F}_k|=e^{-\frac{B}{1024}(\log L)^2}$ and
$\dist({\cal F}_k,{\cal F}_\lambda)\geq 4e^{-\frac{B}{512}(\log L)^2}$, $k\not = \lambda$.
Since the number of gaps between the ${\cal F}_k$ in $\bar{\Delta}$ is less than
$e^{\frac{B}{1024}(\log L)^2}|\bar{\Delta}|$ and $|\bar{\Delta}|<\frac{d(\varepsilon)}{L}$, 
it follows
from Proposition 2 that
\begin{eqnarray}
\mathbb{P}_\Lambda(\Omega_\Lambda{'''})&\equiv&\mathbb{P}_\Lambda\Le(\omega \in
\Omega_\Lambda \,:\, \sigma(H_b)\cap \bar{\Delta}
\subset \bigcup_{k} {\cal F}_k\Ri)
\nonumber\\ 
&\geq& 1- 16c(B) \|h\|_{\infty}\varepsilon^{-2}V_0
L^4e^{-\frac{B}{512}(\log L)^2}e^{\frac{B}{1024}(\log L)^2}
\frac{d(\varepsilon)}{L}\nonumber\\
&=&1- 16c(B)\|h\|_{\infty}\varepsilon^{-2}V_0 d(\varepsilon)
L^3e^{-\frac{B}{1024}(\log L)^2} \; .
\end{eqnarray}
Now suppose that 
$\psi_\beta$ is an eigenstate of $H_\omega$
corresponding to $E_\beta\in
\bar{\Delta}$.
For a given $\omega \in \Omega_\Lambda^{'''}$ one can show that $E_\beta$ is
necessarly included in one of the fattened intervals $\tilde{{\cal F}}_k\equiv 
{\cal F}_k+e^{-\frac{B}{512}(\log L)^2}$. In order to check this it is
sufficient to adapt the estimates \eqref{pdb} to \eqref{bwn} to the difference of 
projectors \mbox{$\|P(\tilde{{\cal F}}_k)-P_b(\tilde{{\cal F}}_k)\|$}. The main point is to check that with our choice of 
intervals one is allowed to replace the circle $\bar{\Gamma}$ by circles 
$\bar{\Gamma}_k$ centered at the midpoint of ${\cal F}_k$ and of diameter 
$e^{-\frac{B}{1024}(\log L)^2} + 2e^{-\frac{B}{512}(\log L)^2}$. We do not give the details here.
One finds
\begin{equation}\label{star}
\|P(\tilde{{\cal F}}_k)-P_b(\tilde{{\cal F}}_k)\|\leq \varepsilon^{-3} C''(B,V_0)Le^{-\frac{B}{1024}(\log
L)^2}\; .
\end{equation}
Therefore $P(\tilde{{\cal F}}_k)\psi_\beta=\psi_\beta$ for some $k$ and we have
\begin{eqnarray}\label{prm}
J_\beta &=& (\psi_\beta,v_y\psi_\beta)  = (\psi_\beta,v_y
P(\tilde{{\cal F}}_k)\psi_\beta) 
=(P_b(\tilde{{\cal F}}_k)\psi_\beta,v_yP_b(\tilde{{\cal F}}_k)\psi_\beta) \\ 
&+&([P(\tilde{{\cal F}}_k)-P_b(\tilde{{\cal F}}_k)]\psi_\beta,v_yP_b(\tilde{{\cal F}}_k)\psi_\beta) 
+
(\psi_\beta,v_y[P(\tilde{{\cal F}}_k)-P_b(\tilde{{\cal F}}_k)]\psi_\beta) \nonumber \; .
\end{eqnarray}
To estimate the
first term on the right hand side of \eqref{prm}
we use the spectral decomposition in terms of eigenstates of $H_b$,
\begin{equation}
P_b(\tilde{{\cal F}}_k)\psi_\beta=\sum_{E_\tau^b\in \tilde{{\cal F}}_k}
(\psi_\tau^b,\psi_\beta)\psi_\tau^b \; .
\end{equation}
We have
\begin{equation}
(P_b(\tilde{{\cal F}}_k)\psi_\beta,v_yP_b(\tilde{{\cal F}}_k)\psi_\beta) = \sum_{
E_\tau^b,E_\sigma^b\in \tilde{{\cal F}}_k} (\psi_\beta,\psi_\tau^b)(\psi_\sigma^b
,\psi_\beta)
(\psi_{\tau}^b,v_y\psi_\sigma^b).
\end{equation}
From Lemma 2 and Lemma 4 in Appendix B we get
\begin{eqnarray}\label{leo}
|(P_b(\tilde{{\cal F}}_k)\psi_\beta,v_yP_b(\tilde{{\cal F}}_k)\psi_\beta)|
&\leq& (\Tr P_b({\cal F}_k))^2 4L e^{-\frac{B}{1024}(\log L)^2}\nonumber \\
&\leq& 16c(B)^4\varepsilon^{-4}V_0^4 L^9e^{-\frac{B}{1024}(\log L)^2}\; .
\end{eqnarray}
The second term on the right hand side of \eqref{prm} 
is estimated by the Schwarz inequality 
\begin{eqnarray}\label{le1}
& &([P(\tilde{{\cal F}}_k)-P_b(\tilde{{\cal F}}_k)]\psi_\beta,v_yP_b(\tilde{{\cal F}}_k)\psi_\beta)^2
\leq
\|v_yP_b(\tilde{{\cal F}}_k)\psi_\beta\|^2\|P(\tilde{{\cal F}}_k)-P_b(\tilde{{\cal F}}_k)\|^2
\nonumber \\
&\leq&2
(P_b(\tilde{{\cal F}}_k)\psi_\beta,(H_b-V_\omega)P_b(\tilde{{\cal F}}_k)\psi_\beta)\|P(\tilde{{\cal F}}_k)-P_b(\tilde{{\cal F}}_k)\|^2 
\nonumber\\ &\leq &
(B+3V_0)\|P(\tilde{{\cal F}}_k)-P_b(\tilde{{\cal F}}_k)\|^2 \; .
\end{eqnarray}
The third term is treated in a similar way
\begin{eqnarray}\label{le2}
(\psi_\beta,v_y[P(\tilde{{\cal F}}_k)-P_b(\tilde{{\cal F}}_k)]\psi_\beta)^2 &\leq&
\|v_y\psi_\beta\|^2\|P(\tilde{{\cal F}}_k)-P_b(\tilde{{\cal F}}_k)\|^2 \nonumber
\\
&\leq&2
(\psi_\beta,(H_   \omega-V_\omega)\psi_\beta)\|P(\tilde{{\cal F}}_k)-P_b(\tilde{{\cal F}}_k)\|^2 
\nonumber\\ 
&\leq&
(B+3V_0)\|P(\tilde{{\cal F}}_k)-P_b(\tilde{{\cal F}}_k)\|^2 \; .
\end{eqnarray}
The last estimate \eqref{rt4} of Theorem 1 then follows from \eqref{star}, \eqref{leo},
\eqref{le1}
and \eqref{le2}.
\end{proof}

\emph{Remark}. The set $\hat{\Omega}_\Lambda$ in Theorem 1 may be taken equal to
$\Omega_\Lambda^{'}\cap\Omega_\Lambda{''}\cap\Omega_\Lambda^{'''}$. 
This set has a probability larger than
$1-3L^{-s}$ with $s=\min(\theta,p-6)$.

\appendix

\section{Resolvent of the Landau Hamiltonian}

The  kernel $R_0(\bold{x},\bold{x}';z)$ of the resolvent $R_0(z)=(z-H_0)^{-1}$
with periodic boundary conditions along $y$ can be expressed in term of the  kernel
$R_0^\infty(\bold{x},\bold{x}';z)$ of the resolvent of the two
dimensional Landau Hamiltonian defined on the whole plane $\R^2$.
Since the spectrum and the eigenfunctions of $H_0$ are 
exactly known, by writing
down the spectral decomposition of $R_0(\bold{x},\bold{x}';z)$ and applying the Poisson
summation formula we get for $z \in  \rho(H_0)$
\begin{equation}\label{P}
R_0(\bold{x},\bold{x}';z) =\sum_{m\in \Z} R_0^\infty(x\,y-mL,x'\,y';z) \; .
\end{equation}
The formula for $R_0^\infty(\bold{x},\bold{x}';z)$ is (see for example \cite{DMP4}) 
\begin{equation}\label{A1}
R_0^\infty(\bold{x},\bold{x}';z) = \frac{B}{2\pi}\Gamma
(\alpha_z)U\Le(\alpha_z,1;\frac{B}{2}|\bold{x}-\bold{x}'|^2\Ri)e^{-\frac{B}{4}|\bold{x}-\bold{x}'|^2}
M(\bold{x},\bold{x}')
\end{equation}
where $\alpha_z=(\frac{1}{2}-\frac{z}{B})$ and 
\begin{equation}
M(\bold{x},\bold{x}')=\exp\Le(\frac{i}{2}B(x+x')(y-y')\Ri)
\end{equation}
is the phase factor in the Landau gauge.
In \eqref{A1} the Landau levels appear as simple poles of the Euler
$\Gamma$ function and
$U(-\lambda,b;\rho)$ is the logarithmic 
solution of the Kummer equation (see eqns. (13.1.1) and (13.1.6) of \cite{AS})
\begin{equation}
\rho\frac{d^2U}{d\rho^2}+(b-\rho)\frac{dU}{d\rho} +\lambda\rho=0 \; .
\end{equation}

\begin{lem}\label{lem3} 
If $|{\cal I}m \,z|\leq 1$, ${\cal R}e \,z \in
\,\Le]\frac{1}{2}B,\frac{3}{2}B\Ri[$ and 
\mbox{$\frac{B}{2}|x-x'|^2>1$} then, for $L$ large enough, there exists
 $C_n(z,B)$, $n=0,1$ independent of $L$ such that
\begin{eqnarray}\label{est1}
|\partial_x^n R_0(\bold{x},\bold{x}';z)| &\leq&  C_n(z,B)
e^{-\frac{B}{8}(x-x')^2}
\end{eqnarray}
where $C_n(z,B)=
C_n B^{1+\frac{n}{2}}\dist(z,\sigma(H_0))^{-1}$ with $C_n$
a numerical positive constant. 
\end{lem}

For our purposes we need only decay in the $x$ direction as provided by the lemma but in
fact there is also a Gaussian decay in the $y$ direction as long as
$|y-y'|<\frac{L}{2}$. One can also prove similar estimates when ${\cal R}e \, z$ 
is between higher Landau levels but the constant is not uniform with respect to
 $\nu$. Finally we
point out that this estimate does not hold for $\frac{B}{2}|\bold{x}-\bold{x}'|^2<1$
because of the  logarithmic singularity in the Kummer function for $\rho\to 0$
(see also Appendix C).

\begin{proof} 
The proof relies on the estimate (6.10) of \cite{DMP4} which we state here for
convenience. For $\lambda=x+iy$, $N-1<x<N$ ($N\geq 1$), $b\in \N$ and $\rho>1$
\begin{eqnarray}
|U(-\lambda,b;\rho)|&\leq&2^{b+N-1}\rho^x(b+N+|y|)^N\frac{|\Gamma(-x)|}{|\Gamma(-\lambda)|}
\nonumber\\
&+&e^{-(\rho-2)}(\rho+1+|y|)^N\frac{(b+N)!}{|\Gamma(N-\lambda)|} \; .
\end{eqnarray}
Using this estimate for $N=1$, $|y|<1$ and $b=n$ together with
$\Gamma(1-\lambda)=-\lambda\Gamma(-\lambda)$ we have ($C_n'$ a numerical constant)
\begin{equation}\label{bb}
|\Gamma(-\lambda)||U(-\lambda,n+1;\rho)|\leq C_n' \rho
\Le\{\Gamma(-x)+|\lambda|^{-1}\Ri\} \; .
\end{equation}
From \eqref{bb} 
if $|{\cal I}m \,z|\leq 1$, ${\cal R}e \, z \in  ]\frac{1}{2}B,\frac{3}{2}B[$
and $\frac{B}{2}|\bold{x}-\bold{x}'|^2>1$ we deduce the estimate ($C_n''$ a numerical constant)
\begin{equation}\label{EHG}
|\Gamma(\alpha_z)U\Le(\alpha_z,n+1;\frac{B}{2}|\bold{x}-\bold{x}'|^2\Ri)| \leq
B C_n''\dist(z,\sigma(H_0))^{-1}|\bold{x}-\bold{x}'|^{2} \; . 
\end{equation}
From \eqref{EHG} for $n=0$ and \eqref{P} we get
\begin{equation}
|R_0(\bold{x},\bold{x}';z)|\leq 2BC''_0\dist(z,\sigma(H_0))^{-1}e^{-\frac{B}{8}(x-x')^2} 
\sum_{m\in \Z} e^{-\frac{B}{8}(y-y'-mL)^2} \;  
\end{equation}
since $|y-y'|<L$ the last sum can be bounded by a constant, which yields  
\eqref{est1} for $n=0$.

To estimate the first derivative it is convenient to use
 the relation \cite{AS}
\begin{equation}\label{ASK}
\frac{\textrm{d}U(-\lambda,1;\rho)}{\textrm{d}\rho}=U(-\lambda,1;\rho)-U(-\lambda,2;\rho)
\end{equation} 
which yields
\begin{eqnarray}\label{CR}
\partial_x R_0^\infty(\bold{x},\bold{x}';z)
&=&\frac{B}{2}\Le[(x-x')+i(y-y')\Ri]R^\infty_0(\bold{x},\bold{x}';z)\\
&-&  B(x-x')\frac{B}{2\pi}
\Gamma(\alpha_z)U\Le(\alpha_z,2;\frac{B}{2}|\bold{x}-\bold{x}'|^{2}\Ri)
e^{-\frac{B}{4}|\bold{x}-\bold{x}'|^{2}}M(\bold{x},\bold{x}')\nonumber \; .
\end{eqnarray}
Using \eqref{EHG} to bound the two terms on the right hand side of 
\eqref{CR} we get
\begin{equation}\label{di}
|\partial_x R_0^\infty(x\, y,x'\, y'-mL;z)|\leq B^{\frac{3}{2}}
C_1''\dist(z,\sigma(H_0))^{-1}
e^{-\frac{B}{8}\Le[(x-x')^2+(y-y'-mL)^2\Ri]}
\end{equation}
the result \eqref{est1} for $n=1$ then follows from \eqref{di} and \eqref{P}.
\end{proof}

\section{Bounds on the Number of Eigenvalues in Small Intervals}

We first prove a deterministic Lemma on the maximal number of eigenvalues of 
$H_b$ belonging to energy intervals $I$ contained in $\Delta_\varepsilon$.
Then we sketch the proof of Proposition 2.
 The ideas in this appendix come
 from the method used by Combes and Hislop to obtain the Wegner estimate
which gives the expected number
of eigenvalues in $I$. Since Lemma 4 does not appear in \cite{CH} and we need to adapt
the technique to our geometry we give some details below. 

 We begin with some preliminary observations
 on the  kernel
$P_0(\bold{x},\bold{x}')$ of the projector onto the first Landau 
level with periodic boundary conditions along $y$. Using the spectral decomposition and the Poisson
summation formula one gets
\begin{equation}\label{pk}
P_0(x \, y,x' \, y') = \sum_{m\in \Z} P_0^\infty(x\, y-mL, x' \, y') 
\end{equation}
where
\begin{equation}
P_0^\infty(\bold{x},\bold{x}') = \frac{B}{2\pi}e^{-\frac{B}{4}|\bold{x}-\bold{x}'|^2}
e^{i\frac{B}{2}(x+x')(y-y')}
\end{equation}
is the projector on the first Landau level for the infinite plane. The above formula 
can also be obtained
by computing the residues of the poles of the $\Gamma$ function.
We observe that $V_{\bold{i}}^{1/2}P_0V_{\bold{j}}^{1/2}$ is trace class. Indeed it is the product of
two Hilbert-Schmidt operators $V_{\bold{i}}^{1/2}P_0$ and $P_0V_{\bold{j}}^{1/2}$ and from the expression
of the kernel \eqref{pk} it is easily seen that ($c(B)$ a constant independent of $L$)
\begin{equation}\label{tn}
\|V_{\bold{i}}^{1/2}P_0V_{\bold{j}}^{1/2}\|_1\leq \|V_{\bold{i}}^{1/2}P_0\|_2
\|P_0V_{\bold{j}}^{1/2}\|_2\leq c(B)V_0 \; .
\end{equation}

\begin{lem}\label{lem12}
Let $I$ be any interval contained in $\Delta_\varepsilon$ and $P_b(I)$ the
eigenprojector associated to $H_b$. Then
\begin{equation}\label{numeig}
\Tr P_b(I)\leq 2\varepsilon^{-2}c(B)^2V_0^2 L^4 \; .
\end{equation}
\end{lem}

\begin{proof}
Let $Q_0=1-P_0$ and $E$
 the middle point of $I$.
Using $Q_0(H_0-E)Q_0\geq 0$ and $Q_0R_0(E)Q_0\leq(B-V_0)^{-1}Q_0$ we can write
\begin{eqnarray}
P_b(I) Q_0 P_b(I) &=& P_b(I) Q_0(H_0-E)^{1/2}R_0(E)(H_0-E)^{1/2}Q_0
P_b(I)\\
&\leq& (B-V_0)^{-1}P_b(I) (H_0-E)Q_0P_b(I) \nonumber\\
&\leq& (B-V_0)^{-1} \Le[P_b(I) (H_b-E)Q_0P_b(I) -
P_b(I) V_\omega Q_0 P_b(I)\Ri] \nonumber
\end{eqnarray}
and thus from $\|P_b(I) (H_b-E)\|\leq \frac{|I|}{2}$, we get
\begin{equation}\label{pqp}
\|P_b(I) Q_0 P_b(I)\| \leq (B-V_0)^{-1}\Le(\frac{|I|}{2}+ V_0\Ri) \leq
\frac{3V_0}{2(B-V_0)}\leq\frac{1}{2} \; .
\end{equation}
In the last inequality we have assumed that $B\geq 4V_0$.
Using $\Tr P_b(I)=\Tr P_b(I) P_0 P_b(I)+\Tr P_b(I) Q_0 P_b(I)$,
$\Tr P_b(I) Q_0 P_b(I) \leq \|P_b(I) Q_0 P_b(I)\| \Tr P_b(I)$, and
\eqref{pqp} we obtain 
\begin{equation}\label{eq1}
\Tr P_b(I) \leq 2\Tr P_b(I) P_0 P_b(I) = 2\Tr P_0 P_b(I) P_0\; .
\end{equation}
Now, from
\begin{equation}
\dist(I,\frac{B}{2})^2P_b(I)^2 \leq
\Le(P_b(I)(H_b-\frac{B}{2})P_b(I)\Ri)^2
\end{equation}
it follows that
\begin{eqnarray}\label{eqq2}
\Tr P_0 P_b(I) P_0 &\leq&
\varepsilon^{-2}\Tr (P_0
P_b(I)(H_b-\frac{B}{2})P_b(I)(H_b-\frac{B}{2})P_b(I)
P_0)
\nonumber\\
&=& \varepsilon^{-2}\Tr(P_0 V_\omega P_b(I) V_\omega P_0) \label{eq2}
\leq \varepsilon^{-2}\|P_0 V_\omega\|_{2}\| V_\omega P_0\|_{2}
\end{eqnarray}
each Hilbert-Schmidt norm in \eqref{eqq2} is bounded by $c(B)V_0L^2$. This observation 
together with \eqref{eq1} gives the result of the lemma.
\end{proof}

Let us now sketch the proof of Proposition 2.

\begin{proof}[Proof of Proposition \ref{prop3}]
Let $E\in\Delta_\varepsilon$ and $I=[E-\delta,E+\delta]$ for $\delta$ small enough
(we require that $I$ is contained in $\Delta_\varepsilon$).
By the Chebyshev inequality we have
\begin{eqnarray}
\mathbb{P}_\Lambda\Le(\dist(\sigma(H_b),E)<\delta\Ri) = 
\mathbb{P}_\Lambda\Le(\Tr P_b(I)\geq 1\Ri)
\leq
\mathbb{E}_\Lambda(\Tr P_b(I))
\end{eqnarray}
where $\mathbb{E}_\Lambda$ is the expectation with respect to the random variables in $\Lambda$.
To estimate it we use an intermediate inequality of the previous proof
\begin{equation}\label{iin}
\mathbb{E}_\Lambda(\Tr P_b(I))
\leq 2\varepsilon^{-2}\mathbb{E}_\Lambda(
\Tr P_0 V_\omega P_b(I) V_\omega P_0) \; .
\end{equation}
Writing  $V_{\omega,\Lambda}=\sum_{\bold{i}\in \Lambda} X_{\bold{i}}(\omega)V_{\bold{i}}$
\begin{eqnarray}\label{33}
\Tr P_0 V_\omega P_b(I) V_\omega P_0 &=&
\sum_{\bold{i},\bold{j} \in \Lambda^2} X_{\bold{i}}(\omega)X_{\bold{j}}(\omega) \Tr P_0
V_{\bold{i}}P_b(I) V_{\bold{j}} P_0 \label{eq3} \\
&=&\sum_{\bold{i},\bold{j} \in \Lambda^2} X_{\bold{i}}(\omega)X_{\bold{j}}(\omega)
\Tr V_{\bold{j}}^{1/2}P_0
V_{\bold{i}}^{1/2} V_{\bold{i}}^{1/2} P_b(I) V_{\bold{j}}^{1/2}\; .   \nonumber
\end{eqnarray}
Since $V_{\bold{j}}^{1/2}P_0V_{\bold{i}}^{1/2}$ is trace 
class we can introduce the singular value decomposition
\begin{equation}
V_{\bold{j}}^{1/2}P_0V_{\bold{i}}^{1/2} =\sum_{n=0}^\infty\mu_n(\psi_n,.)\phi_n
\end{equation}
where $\sum_{n=0}^\infty\mu_n=\|V_{\bold{j}}^{1/2}P_0V_{\bold{i}}^{1/2}\|_1$. Then
\begin{eqnarray}\label{44}
& &\Tr V_{\bold{j}}^{1/2}P_0
V_{\bold{i}}^{1/2} V_{\bold{i}}^{1/2}  P_b(I) V_{\bold{j}}^{1/2} =\sum_{n=0}^\infty\mu_n
(\phi_n,V_{\bold{i}}^{1/2} P_b(I) V_{\bold{j}}^{1/2}\psi_n)\nonumber
\\
&\leq& \sum_{n=0}^\infty\mu_n(\phi_n, V_{\bold{i}}^{1/2} P_b(I) V_{\bold{i}}^{1/2}\phi_n)^{1/2}
(\psi_n, V_{\bold{j}}^{1/2} P_b(I) V_{\bold{j}}^{1/2}\psi_n)^{1/2} \nonumber\\
&\leq& 
\frac{1}{2}\sum_{n=0}^\infty \mu_n  \Le\{(\phi_n,V_{\bold{i}}^{1/2} P_b(I) V_{\bold{i}}^{1/2}\phi_n)+
(\psi_n,V_{\bold{j}}^{1/2} P_b(I) V_{\bold{j}}^{1/2}\psi_n)\Ri\}  \; .
\end{eqnarray}
An application of the spectral averaging theorem of \cite{CH} shows that 
\begin{equation}\label{spav}
\mathbb{E}_\Lambda((\psi_n,V_{\bold{j}}^{1/2} P_b(I) V_{\bold{j}}^{1/2}\psi_n))
\leq \|h\|_{\infty}2\delta
\end{equation}
as well as for the term with $\bold{i}$ replacing $\bold{j}$ and $\phi_n$ replacing $\psi_n$.
Combining \eqref{iin}, \eqref{44}, \eqref{spav} and \eqref{eq3} we get
\begin{equation}
\mathbb{E}_\Lambda(\Tr P_b(I))
\leq 
4\|h\|_{\infty}\delta\varepsilon^{-2}
\sum_{\bold{i},\bold{j} \in \Lambda^2}\|V_{\bold{j}}^{1/2} P_0 V_{\bold{i}}^{1/2}\|_1
\leq
4\|h\|_{\infty}\delta\varepsilon^{-2}c(B) V_0L^4 \; .
\end{equation}
\end{proof}

\section{Estimate on the Eigenfunction of $H_b$}

In this section we prove Gaussian decay of the eigenfunction $\psi_\beta^b$ and its
$y-$derivative outside the support of the random potential $V_\omega$.
From the eigenvalue equation $(H_0+V_\omega)\psi_\beta^b=E_\beta^b\psi_\beta^b$
we get 
\begin{equation}
\psi_\beta^b=R_0(E_\beta^b)V_\omega\psi_\beta^b \; .
\end{equation}
Thus
\begin{eqnarray}\label{psi}
|\psi_\beta^b(\bold{x})| &\leq& \int_{\R\times I_p}
|R_0(\bold{x},\bold{x}';E_\beta^b)V_\omega(\bold{x}')\psi_\beta^b(\bold{x}')| \D
\bold{x}'
\nonumber \\
&\leq& V_0 \Le\{\int_{\supp V_\omega}|R_0(\bold{x},\bold{x}';E_\beta^b)|^2 \D \bold{x}'
\Ri\}^{1/2}\; ,
\end{eqnarray}
and
\begin{equation}\label{ypsi}
|\partial_y \psi_\beta^b(\bold{x})| \leq V_0 \sup_{\bold{x}} |\psi_\beta^b(\bold{x})|
\int_{\supp V_\omega} |\partial_y R_0(\bold{x},\bold{x}';E_\beta^b)| \D
\bold{x}'\; .
\end{equation}

We need bounds on the integral kernel $R_0$ and its
$y-$derivative to get an estimate of the eigenfunctions and their $y-$derivative.
From \cite{DMP4} we have ($E\in \Delta_\varepsilon$)
\begin{eqnarray}\label{G}
|R_0^\infty(\bold{x},\bold{x}';E)| &\leq&
C(B)|\Gamma(\alpha_E)|e^{-\frac{B}{8}|\bold{x}-\bold{x}'|^2}\times \nonumber \\
&\times&
\begin{cases}
1 &\textrm{if $\frac{B}{2}|\bold{x}-\bold{x}'|^2>1$}\\
1+\Le|\ln (\frac{B}{2}|\bold{x}-\bold{x}'|^2)\Ri|
&\textrm{if $\frac{B}{2}|\bold{x}-\bold{x}'|^2\leq1$} \; .\\
\end{cases}
\end{eqnarray}
Calculating the $y-$derivative thanks to \eqref{ASK}, and using bounds (6.16) of
\cite{DMP4} we have
\begin{eqnarray}\label{yG}
& &|\partial_y R_0^\infty(\bold{x},\bold{x}';E)| \leq
C'(B)|\Gamma(\alpha_E)|e^{-\frac{B}{8}|\bold{x}-\bold{x}'|^2}\times \nonumber \\
&\times&\begin{cases}
1+|x| &\textrm{if $\frac{B}{2}|\bold{x}-\bold{x}'|^2>1$}\\
\Le(1+\Le|\ln
(\frac{B}{2}|\bold{x}-\bold{x}'|^2)\Ri|\Ri)(1+|x|+|\bold{x}-\bold{x}'|^{-1})
&\textrm{if $\frac{B}{2}|\bold{x}-\bold{x}'|^2\leq1$} \; .\\
\end{cases}
\end{eqnarray}

With the help of \eqref{G} and \eqref{yG} we can see that 
for $L$ large enough
\begin{equation}\label{LL1}
|\psi_\beta^b(\bold{x})| \leq
C(B)\varepsilon^{-1}V_0 L \times 
\begin{cases}
e^{-\frac{B}{8}(x-\frac{L}{2}+\log L)^2} &\textrm{if $x\not \in
\Le[-\frac{L}{2},\frac{L}{2}\Ri]$}\\
\ln(BL^2) &\textrm{if $x \in
\Le[-\frac{L}{2},\frac{L}{2}\Ri]$} \; .\\
\end{cases}
\end{equation}
and 
\begin{equation}\label{LL2}
|\partial_y \psi_\beta^b(\bold{x})| \leq
C'(B)\varepsilon^{-2}V_0^2 L^2 \times 
\begin{cases}
e^{-\frac{B}{8}(x-\frac{L}{2}+\log L)^2}(1+|x|)  &\textrm{if $x\not \in
\Le[-\frac{L}{2},\frac{L}{2}\Ri]$}\\
L(\ln(BL^2)^2(1+|x|) &\textrm{if $x \in
\Le[-\frac{L}{2},\frac{L}{2}\Ri]$} \; .\\
\end{cases}
\end{equation}
Indeed, for $|m|>1$ $\frac{B}{2}[(x-x')^2+(y-y'-mL)^2]>1$ thus we have
\begin{equation}
|R_0(\bold{x},\bold{x}';E_\beta^b)|\leq \tilde{C}(B) \varepsilon^{-1}
e^{-\frac{B}{8}(x-x')^2} +\sum_{|m|\leq 1} |R_0^\infty(x\, y, x'\,
y'-mL;E_\beta^b)| \; .
\end{equation}
If $x\not \in\Le[-\frac{L}{2},\frac{L}{2}\Ri]$ since $\bold{x}'\in \supp
V_\omega$ the terms $|m|\leq 1$ have also a Gaussian bound and
\begin{equation}
|R_0(\bold{x},\bold{x}';E_\beta^b)|\leq \tilde{C}'(B) \varepsilon^{-1}
e^{-\frac{B}{8}(x-x')^2} \; .
\end{equation}
Replacing this bound in \eqref{psi} we get the Gaussian decay in \eqref{LL1}
On the other hand 
if $x \in\Le[-\frac{L}{2},\frac{L}{2}\Ri]$ we can use the logarithmic bounds for
the terms $|m|\leq 1$ and we remark they are integrable and bounded by
$L^2\ln(BL^2)$.
The same arguments hold for the $y-$derivative.

\section*{Acknowledgements} N.M. wishes to thank J. Fr\" ohlich and E. Mourre
for helpful discussions and F. Bentosela for suggesting the 
use of the decoupling formula.
The work of C.F. was supported by a grant from the Fonds National Suisse
de la Recherche Scientifique No. 20 - 55654.98.

\end{document}